\documentclass[conference]{IEEEtran}
\IEEEoverridecommandlockouts

\usepackage[frozencache=true,cachedir=minted-cache]{minted} 

\usepackage{cite}
\usepackage{amsmath,amssymb,amsfonts}
\usepackage{algorithmic}
\usepackage{graphicx}
\usepackage{textcomp}
\usepackage{xcolor}
\usepackage{makecell}
\usepackage{multirow}
\usepackage{caption}
\usepackage{subcaption}
\usepackage{hyperref}
\usepackage{footnote}
\usepackage{footmisc}
\usepackage{soul}
\usepackage{setspace} 
\usepackage{tabularx}

\setminted{fontsize=\small,baselinestretch=1}
\newcommand{\ben}[1]{#1}

\newenvironment{rednote}{\par\color{red}}{\par}
\newcommand{\wei}[1]{{\color{blue}Wei:~[#1]}}

\newcommand{\mycomment}[1]{}

\newcommand{\devign}{Devign\ }
\newcommand{\msr}{MSR\ }
\newcommand{\xxx}{\textcolor{red}{XXX}\ }
\newcommand{\wip}[1]{\textcolor{red}{#1}}
\newcommand{\needcite}{\textcolor{red}{(citation needed)}\ }
\newcommand{\dataurl}{https://doi.org/10.6084/m9.figshare.20791240}

\newcommand{\symbfootnote}[1]{%
\let\oldthefootnote=\thefootnote%
\setcounter{footnote}{0}%
\renewcommand{\thefootnote}{\fnsymbol{footnote}}%
\footnote{#1}%
\let\thefootnote=\oldthefootnote%
}

\def\BibTeX{{\rm B\kern-.05em{\sc i\kern-.025em b}\kern-.08em
    T\kern-.1667em\lower.7ex\hbox{E}\kern-.125emX}}

\begin{document}

\title{An Empirical Study of Deep Learning Models\\ for Vulnerability Detection
}

\author{
\IEEEauthorblockN{Benjamin Steenhoek}
\IEEEauthorblockA{
Iowa State University\\
Ames, Iowa, USA \\
benjis@iastate.edu}
  \and
\IEEEauthorblockN{Md Mahbubur Rahman}
\IEEEauthorblockA{
Iowa State University\\
Ames, Iowa, USA \\
mdrahman@iastate.edu}
  \and
\IEEEauthorblockN{Richard Jiles}
\IEEEauthorblockA{
Iowa State University\\
Ames, Iowa, USA \\
rdjiles@iastate.edu}
  \and
\IEEEauthorblockN{Wei Le}
\IEEEauthorblockA{
Iowa State University\\
Ames, Iowa, USA \\
weile@iastate.edu}
}


\maketitle

\begin{abstract}
Deep learning (DL) models of code have recently reported great progress for vulnerability detection. In some cases, DL-based models have outperformed static analysis tools. Although many great models have been proposed, we do not yet have a good understanding of these models. This limits the further advancement of model robustness, debugging, and deployment for the vulnerability detection. In this paper, we surveyed and reproduced 9 state-of-the-art (SOTA) deep learning models on 2 widely used vulnerability detection datasets: Devign and MSR. We investigated 6 research questions in three areas, namely {\it model capabilities}, {\it training data}, and {\it model interpretation}. We experimentally demonstrated the variability between different runs of a model and the low agreement among different models' outputs. We investigated models trained for specific types of vulnerabilities compared to a model that is trained on all the vulnerabilities at once.
We explored the types of programs DL may consider ``hard'' to handle. We investigated the relations of training data sizes and training data composition with model performance. Finally, we studied model interpretations and analyzed important features that the models used to make predictions. We believe that our findings can help better understand model results, provide guidance on preparing training data, and improve the robustness of the models. All of our datasets, code, and results are available at \url{\dataurl}.
\end{abstract}


\mycomment{
{\color{red}

We studied cross-sections of the Devign and MSR datasets to understand the effects of attributes like the bug type, dataset size, and project composition in the training and test datasets.
We also studied the models' variations over random seeds and developed approaches based on logistic regression (LR) and interpretability to understand the deep learning models' inputs and behavior.
Among other things, we found that the models could not generalize to unseen bug types and projects,
and that 
certain code features were difficult for the models to handle.
Our findings motivate future work in improving vulnerability detection models and datasets, and provide a starting point for further directions of empirical study.

}
}


\section{Introduction}
Deep learning vulnerability detection tools have achieved promising results in recent years. The state-of-the-art (SOTA) models reported 0.9 F1 score~\cite{linevul,hgvul} and outperformed static analyzers~\cite{ding2022velvet,mvd}.  The results are exciting in that deep learning may bring in transformative changes for software assurance. Thus, industry companies such as IBM, Google and Amazon are very interested and have invested heavily to develop such tools and datasets~\cite{codexglue,codenet,alphacode,d2a_paper}.

Although promising, deep learning vulnerability detection has not yet reached the level of computer vision and natural language processing. Most of our research focuses on trying a new emerging deep learning model and making it work for a dataset like the Devign or MSR dataset~\cite{bigvul,zhou_devign_2019,codexglue}. However, we know little about the model itself, e.g., what type of programs the model can/cannot handle well, whether we should build \ben{models} for each vulnerability type or we should build one model for all \ben{vulnerability types}, what is a good training dataset, and what information the model has used to make the decisions. Knowing the answers to these questions can help us better develop, debug, and apply the models in practice. But considering the black-box nature of deep learning, these questions are very hard to answer. This paper does not mean to provide a complete solution for these questions but is an exploration towards these goals.


In this paper, we surveyed and reproduced a collection of SOTA deep learning vulnerability detection models, and constructed research questions and studies to understand these models, with the goal of distilling lessons and guidelines for better designing and debugging future models. To the best of our knowledge, this is the first paper that systematically investigated and compared a variety of SOTA deep learning models. In the past, Chakraborty et al.~\cite{are_we_there_yet} have explored four existing models such as VulDeePecker~\cite{vuldeepecker}, SySeVR~\cite{sysevr} and Devign~\cite{zhou_devign_2019} and pointed out that the models trained with synthetic data reported low accuracies on real-world test set, and the models used spurious features like variable names to make the predictions.

We constructed our research questions and classified them into three areas, namely {\it model capabilities}, {\it training data}, and {\it model interpretation}. Specifically, our first goal is to understand the capabilities of deep learning for handling vulnerability detection problems, especially regarding the following research questions:


\begin{itemize}
    \item {\bf RQ1} Do models agree on the vulnerability detection results? What are the variabilities across different runs of a model and across different models?
     \item {\bf RQ2} Are certain types of vulnerabilities easier to detect? Should we build models for each type of vulnerabilities or should we build one model that can detect all the vulnerabilities?
    \item {\bf RQ3} Are programs with certain code features harder to be predicted correctly by current models, and if so, what are those code features?
\end{itemize}

Our second study focuses on training data. We aim to understand whether and how the training data size and project composition can affect the model performance.  Specifically, we constructed the following research questions:

\begin{itemize}
    \item {\bf RQ4} Can increasing the dataset size help improve the model performance for vulnerability detection?
    \item {\bf RQ5} How does the project composition in the training dataset affect the performance of the models?
    
 \end{itemize}
 
Finally, our third investigation area is model intepretation. We used SOTA model explanation tools to investigate: 
 
 \begin{itemize}
    \item {\bf RQ6} What source code information the models used for prediction?  Do the models agree on the important features?
 \end{itemize}

To answer the research questions, we surveyed the SOTA deep learning models and successfully reproduced 11 models on their original datasets (see Section \ref{sec:survey}). These models used different deep learning architectures such as GNN, RNN, LSTM, CNN, and Transformers. To compare the models, we managed to make 9 models work with the Devign and MSR, two popular datasets. We selected the two datasets because (1) both of the datasets contain real-world projects and vulnerabilities; (2) the majority of models are evaluated and tuned with the Devign dataset in their papers; and (3) the MSR dataset contains 310 projects and its data have annotations on vulnerability types, which are needed to study our RQs. We discovered the findings for our 6 RQs with carefully designed experiments (Section \ref{sec:rq}) and considerations of the threats (Section \ref{sec:threats}). In summary, our research contributions include:


\begin{enumerate}
 \item We conducted a comprehensive survey for the deep learning vulnerability detection models.
 \item We delivered a reproduction package, consisting of the trained models and datasets for 11 SOTA deep learning frameworks with various study settings; 
 \item We designed 6 RQs to understand model capabilities, training data and model interpretation; 
 \item We constructed the studies and experimentally obtained the results for the RQs; and  
 \item We prepared interesting examples and data for further studying model interpretability.
 

 \end{enumerate}

 



\section{A Survey of Models and their reproduction} \label{sec:survey}
To collect the SOTA deep learning models, we studied the papers from 2018 to 2022 and also used Microsoft's CodeXGLUE leaderboard \footnote{\url{https://microsoft.github.io/CodeXGLUE}} and IBM's Defect detection D2A leaderboard \footnote{\url{https://ibm.github.io/D2A}}.
We worked with all the open-source models we can find, and successfully reproduced 11 models. The complete list of models and the reasons we failed to reproduce some models are given in our data replication package.




As shown in Table~\ref{survey}, the reproduced models cover a variety of deep learning architectures. Devign\cite{zhou_devign_2019} and ReVeal\cite{are_we_there_yet} used GNN on {\it property graphs}~\cite{zhou_devign_2019} that integrate control flow, data dependencies and AST. ReGVD\cite{regvd} used GNN on tokens. Code2Vec used {\it multilayer perceptron (MLP)} on AST. VulDeeLocator\cite{vuldeelocator} and SySeVR\cite{sysevr} are based the sequence models of RNN and Bi-LSTMs. Recent deep learning detection used pre-trained transformers, including CodeBERT\cite{codebert}, VulBERTa-CNN\cite{vulberta}, VulBERTa-MLP, PLBART\cite{plbart} and LineVul\cite{linevul} 



\begin{table}[htbp]
\caption{11 Reproduced Models\vspace{-0.2cm}}\label{survey}
\begin{center}
\resizebox{\columnwidth}{!}{
\begin{tabular}{|l||c|l|l|}
\hline
Model    &  Year & Architecture & Dataset \\\hline\hline
Devign \ben{\cite{zhou_devign_2019}}   &  2019 & \makecell[l]{ GNN,\\property graph}  & Devign  \\\hline
ReVeal \ben{\cite{are_we_there_yet}}   &  2021 & \makecell[l]{ GNN,\\property graph}  & Devign, ReVeal \\\hline
ReGVD \ben{\cite{regvd}}    &  2022 & GNN, token           & Devign \\\hline\hline
CodeBERT \ben{\cite{codebert}} &  2020 & \makecell[l]{Transformer}   & Devign \\\hline

\multirow{2}{*}{\ben{VulBERTa-CNN \cite{vulberta}}} & \multirow{2}{*}{\ben{2021}} & \multirow{2}{*}{\ben{Transformer, CNN}} & \multirow[t]{4}{2.2cm}{\ben{VulDeePecker, Draper, ReVeal $\mu$VulDeePecker, Devign, D2A} } \\ &  &  & \\ \cline{1-3}
\multirow{2}{*}{\ben{VulBERTa-MLP} \ben{\cite{vulberta}}} & \multirow{2}{*}{\ben{2021}} & \multirow{2}{*}{\ben{Transformer, MLP}} & \\ &  &  & \\ \hline

PLBART \ben{\cite{plbart}}   &  2021 & Transformer          & Devign \\\hline
LineVul \ben{\cite{linevul}}  & 2022      & Transformer          & MSR \\\hline\hline
Code2Vec \ben{\cite{code2vec}} & 2021  & MLP, AST              & Devign \\\hline\hline
SeSyVR \ben{\cite{sysevr}} & 2018 & RNN & SARD, NVD \\\hline 
VulDeeLocator \ben{\cite{vuldeelocator}} & 2020 & Bi-LSTM & SARD, NVD\\\hline 
\end{tabular}
}
\end{center}
\end{table}
For our RQs, we used the \ben{Devign \cite{zhou_devign_2019}} and \ben{MSR \cite{bigvul}} datasets. We studied the datasets used in these 11 models in their original papers, shown under {\it Dataset} in Table~\ref{survey}.  We found that the Devign dataset has been evaluated and tuned in 8 out of 11 models. It is a {\it balanced} dataset consisting of \ben{roughly} the same number of vulnerable and non-vulnerable examples, a total of 27,318 data points (each example is also called an data point).  LineVul worked with the MSR dataset, which is a more recently available dataset. It is an {\it imbalanced dataset}, consisting of 10,900 vulnerable examples and 177,736 non-vulnerable examples. \ben{These examples are labeled with their source projects and their Common Weakness Enumeration entries (CWE), which indicates the types of the vulnerabilities}.
We leverage such traits of the datasets for some of our RQs.



We reproduced the results for the models using their original datasets and settings, shown in Table~\ref{tab:reproduction}. The columns of {\it A}, {\it P}, {\it R} and {\it F} represents the commonly used metrics in deep learning vulnerability detection, including {\it accuracy}, {\it precision}, {\it recall} and {\it F1}.  Our reproduction results are able to  compute within 2\% difference compared with the original papers in general. The exceptions are ReVeal, for which the authors confirmed that our results fixed a data leakage error in the original paper, and Devign, for which we used the third-party\footnote{\url{https://github.com/saikat107/Devign}} reproduction released by Chakaborthy et al \cite{are_we_there_yet}, as the original Devign code is not open-sourced.


\begin{table}
\centering
\caption{Model reproduction on their original datasets. Reproduction results are reported as the mean of 3 random seeds. '-' indicates that the results for a metric were not reported in their papers. The numbers are in percentage.~\label{tab:reproduction}}
\resizebox{\columnwidth}{!}{
\begin{tabular}{|l||c|c|c|c||c|c|c|c|} 
\hline
\multirow{2}{*}{Model} & \multicolumn{4}{c||}{Paper Results} & \multicolumn{4}{c|}{Our Reproduction}  \\ 
\cline{2-9}
                                                & A   & P   & R   & F           & A'  & P'  & R'  & F'                   \\ 
\hline\hline
Devign                                  & 59 & 54 & 63 & 57         & 56 & 50 & 71 & 59                  \\ 
\hline
ReVeal                                  & 63 & 57 & 75 & 64         & 53 & 48 & 71 & 56                  \\ 
\hline
ReGVD                               & 63 & -   & -   & -           & 62 & 62 & 46 & 52                  \\ 
\hline
CodeBERT                                & 62 & -   & -   & -           & 64 & 59 & 54 & 55                  \\ 
\hline
VulBERTa-CNN
& 64 & -   & -   & -           & 64 & 60 & 59 & 59                  \\ 
\hline
VulBERTa-MLP
& 65 & -   & -   & -           & 63 & 60 & 58 & 59                  \\ 
\hline
PLBART                                 & 63 & -   & -   & -           & 62 & 58 & 59 & 59                  \\ 
\hline
LineVul                               & -   & 97 & 86 & 91         & 99 & 96 & 88 & 92                  \\ 
\hline\hline
Code2Vec                              & 62 & -   & -   & -           & 59 & 55 & 58 & 57                  \\
\hline
\hline 
SeSyVR & 98 & 90 & 92 & 90 &94 &	88 &	84 &	86\\\hline 
VulDeeLocator  & 99 & 98  & - & 97&98 &	99 &	96 &	98\\\hline 

\end{tabular}

}
\end{table}

To enable the comparisons of the models, we improved the models' implementations to support both Devign and MSR datasets.
When running experiments for the RQs, we excluded {\it VulDeeLocator} and {\it SeSyVR} as they cannot be easily modified for the Devign and MSR datasets.
As a result, we used the rest 9 models for our studies of the RQs.

\section{Research Questions and Findings}\label{sec:rq}
We organized the research questions into three areas, namely {\it model capabilities}, {\it training data}, and {\it model interpretation}. See Sections \ref{sec:model-capabilities} to \ref{sec:model-interpretation} respectively. For each RQ, we present the motivation, study setup and our findings.  





\subsection{Capabilities of Deep Learning Models}\label{sec:model-capabilities}





\noindent{\bf RQ1} Do models agree on the vulnerability detection results? What are the variabilities across different runs of a model and across different models?

\vspace{0.1cm}
\noindent{\bf Motivation:} It is known that deep learning model performance can vary across training runs when using different random seeds. In this RQ, we aim to measure for vulnerability detection, how much such variability actually exists. Additionally, we want to discover how much agreement exists across different deep learning models and across the models with similar architectures.
We hope our findings can inform developers and researchers of the uncertainty which potentially exists behind the numbers reported by such tools.


\vspace{0.1cm}
\noindent{\bf Study Setup:} We trained the models using 3 different random seeds on the same train/valid/test partitions of the Devign dataset. We used this dataset because
almost all the models tuned their hyperparameters on it.
We measured the percentage of {\it stable} inputs----an input that has the same binary label for all 3 random seeds. We then compared the stable inputs across the models to measure their agreement.





\vspace{0.1cm}
\noindent{\bf Findings:}  In~Table \ref{fig:rqa3-stability},  we reported the percentage of stable inputs for the entire dataset, under {\it stable-all}, and for the test dataset, under {\it stable-test}. We also reported the variations of the F1 score across 3 seeds (on test dataset) under {\it stdev-test-F1}.

Our results show that on average 34.9\% test data (30.6\% total data) reported different predictions dependent on the seeds used in training. The GNN models that work on property graph ranked the top 2 variability; especially for ReVeal, for 50\% of the test data, its outputs changed between runs. Code2Vec reported the least variability compared to the GNN and transformer models. Interestingly, we found that unstable inputs are associated with more incorrect predictions----stable inputs had a total of 19\% incorrect predictions across all seeds and unstable had 47\%.

Although many examples reported different predictions between runs, we found that F1 test scores did not change as much, and had a standard deviation of 2.9 on average. That said, for most models, we expect 95\% of performance measurements to be within a range of 5.8\% above or below the mean performance when measured on multiple random seeds.





\begin{table}[h]
\centering
\caption{Variability over 3 random seeds on Devign dataset \label{fig:rqa3-stability}}
\begin{tabular}{|l||c|c|c|}
\hline
Model        & \multicolumn{1}{c|}{stable-all} & \multicolumn{1}{l|}{stable-test} & \multicolumn{1}{c|}{stdev-test-F1} \\ \hline\hline
ReVeal       & 55\% & 50\% & {2.73} \\ \hline
Devign       & 57\% & 55\% & {2.24} \\ \hline
VulBERTa-MLP & 60\% & 58\% & {3.13} \\ \hline
PLBART       & 72\% & 67\% & {1.03} \\ \hline
LineVul      & 72\% & 67\% & {3.46} \\ \hline
CodeBERT     & 72\% & 69\% & {2.78} \\ \hline
VulBERTa-CNN & 74\% & 71\% & {2.60} \\ \hline
ReGVD        & 74\% & 72\% & {7.33} \\ \hline
Code2Vec     & 89\% & 77\% & {0.78} \\ \hline
\end{tabular}
\end{table}


Table \ref{fig:rqa3-agreement} shows that the deep learning models learned diverse classifiers in that only 7\% of the test data (and 7\% total data) are agreed by all the models. The 3 GNN models agreed on 20\% of test examples (and 25\% total), whereas the 3 top performing transformers (LineVul, PLBART, and VulBERTa-CNN) agreed on 34\% test data (and 44\% total). But when we compared all 5 transformer models, only 22\% of test examples (and 29\% total) are agreed. The low agreement among different models implies that when there are no ground truth labels, a {\it differential testing} approach that compares across models as an oracle may have  limited uses.

\begin{table}[h]
\centering
\caption{Agreement across different models\label{fig:rqa3-agreement}}
\begin{tabular}{|l||c|c|}
\hline
Model        & \multicolumn{1}{l|}{agreed-all}  & \multicolumn{1}{l|}{agreed-test} \\ \hline\hline
All 9 models & 7\% & 7\% \\\hline
All 3 GNN models & 25\% & 20\% \\ \hline
Top 3 transformer models
& 44\% & 34\% \\\hline
All 5 transformer models & 29\% & 22\% \\ \hline
\end{tabular}
\end{table}

\noindent{\bf RQ2} 
Are certain types of vulnerabilities easier to detect? Should we build models for each type of vulnerabilities or should we build one model that can detect all  vulnerabilities?

\vspace{0.1cm}

\noindent{\bf Motivation:} In traditional software assurance techniques such as program analysis, we use different algorithms to detect different vulnerabilities. Certain types, e.g., infinite loops, are more difficult to detect than other types, e.g., memory leaks, because one requires to track symbolic values and reasoning about the loops, and the other only needs to check if the memory free is invoked after its allocation. In this RQ, we are interested to learn whether for deep learning vulnerability detectors, it is also true that certain types of vulnerabilities are easier to detect than the others. Considering different types of vulnerabilities have different semantics and root causes, we also want to gain some insights as to whether we should build a model for each type of vulnerability or for vulnerability in general (without separating them into the types), \ben{like most current work has done}.



\vspace{0.1cm}
\noindent{\bf Study Setup}: Here, we study models based on the vulnerability types, and thus we used the MSR dataset. The examples in the MSR dataset are annotated with CWE~\footnote{\url{https://cwe.mitre.org}}. Using these CWE types, we group the vulnerabilities into 5 categories, namely {\it buffer overflow}, {\it value error}, {\it resource error}, {\it input validation error}, and {\it privilege escalation}, shown under {\it Vulnerability Types} in Table~\ref{bugtype}. Our criteria are that (1) each group contains the bugs of similar root causes and semantics, and (2) each group has a sufficiently large dataset for effectively training the models. Column {\it Total} lists the number of the examples collected from the MSR datasets, including all vulnerable and their patched examples that have a CWE annotation.


Specifically, buffer overflow is caused by reading or writing to the memory outside the bounds of the buffer, e.g, CWE-125 \ben{``Out-of-bounds Read''} and CWE-787 \ben{``Out-of-bounds Write''}. The mapping of the complete CWE list to the 5 groups is given in our dataset.  {\it Value error} includes the examples of CWE-190 \ben{``Integer Overflow''}, CWE-369 \ben{``Divide By Zero''} and \ben{CWE-682} \ben{``Incorrect calculation''}. Such errors are caused by propagating incorrect values through data processing or arithmetic operations. {\it Resource error} is caused by incorrectly freeing or using a resource such as memory or a file pointer, and are typically detected using typestate analysis~\cite{typestate_analysis}. The CWE examples include CWE-415 \ben{``Double Free''} and CWE-404 \ben{``Improper Resource Shutdown''}. {\it Input validation error} is caused by using an external input without validating whether it is correct/benign, e.g., CWE-134 \ben{``Use of Externally-Controlled Format String''} and CWE-89 \ben{``Improper Neutralization of Special Elements used in an SQL Command''} (`SQL Injection'). They are often detected using taint analysis \ben{\cite{tripp_taj_2009}}.  Finally, {\it Privilege escalation} is caused by missing proper permission checks and allowing an unauthorized entity to execute privileged commands or view privileged data, such as CWE-264: \ben{``Permissions, Privileges, and Access Controls''}, and CWE-255: \ben{``Credentials Management Errors''}.   

\begin{table}[htbp]
\caption{Five Types of Vulnerabilities}
\begin{center}
\begin{tabular}{|l||c|c|}
\hline
Vulnerability Type             & Total &  CWE examples  \\\hline\hline
Buffer overflow & \ben{37,291} & CWE-125, CWE-787\\\hline
Value error & 15,126 & CWE-190, CWE-369\\\hline
Resource error & 33,748 & CWE-415, CWE-404\\\hline
Input validation error & 25,514 & CWE-134, CWE-89\\\hline
Privilege escalation & 32,749 & CWE-264, CWE-255\\\hline
\end{tabular}
\label{bugtype}
\end{center}
\end{table}

We partitioned the dataset for each bug type into train, valid, and test datasets with 80\%/10\%/10\% ratios. We trained 5 models using the 5 groups of bugs respectively, as well as a {\it Combined Model} trained with all the bug types for comparison. This reflects the real-world scenarios in that a model trained with specific vulnerability type may be more focused, but the combined model can train with more data. We report the \textit{same-bugtype} performance and \textit{cross-bugtype} performances for each model. The same-bugtype performance reports the test F1 score when the training and test data have the same bug type. The cross-bugtype performances report the test F1 scores when the training and test data have different bug types. 

In this experiment, VulBERTa-CNN, VulBERTa-MLP and some models for Code2Vec did not report valid results because they always predicted the same class on the test data.





\begin{figure*}
    \centering
    \includegraphics[width=\textwidth,keepaspectratio]{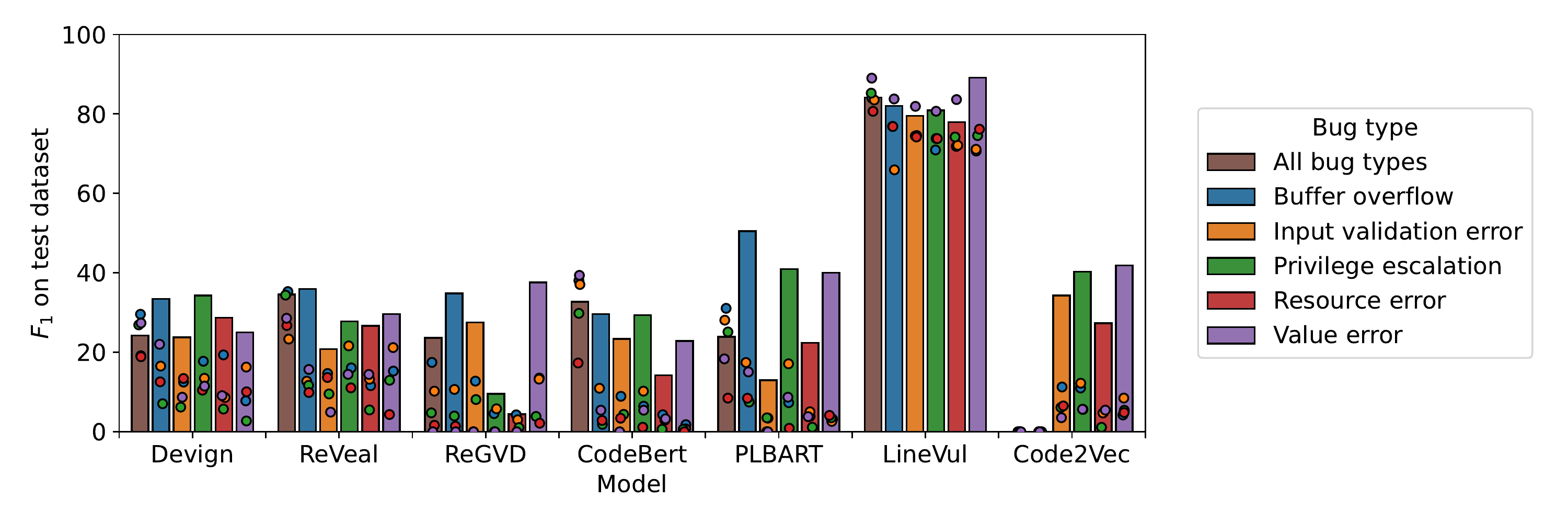}
    \caption{Same-bugtype and cross-bugtype performance. Bars indicates same-bugtype performance; circles indicate cross-bugtype performance for each other bugtype.
    }
    \label{fig:rqa2}
\end{figure*}

\vspace{0.1cm}
\noindent{\bf Findings}: We present out results in Figure~\ref{fig:rqa2}. The bars report the F1 score for the same-bugtype setting, and the circles report the cross-bugtype performances. Each bug type is associated with 4 cross-bugtype test sets, and thus we have four circles for each bar, except that the combined model has five circles, each of which represents running tests for a bug type on the combined models.

Analyzing same-bugtype performance, we found that the models did not always agree on which type of vulnerability is the easiest, and different types of vulnerabilities have achieved the best F1 score in different models. Interestingly, Input validation and Resource errors (the orange and red bars) often reported lower performance than the other types. On the contrary,  Buffer overflows and {Value errors} (the blue and purple bars) often reported better performance compared to other types. In traditional program analysis, these types are harder to detect because they require tracking variable values, and sometimes reasoning about loops.

Devign and ReVeal used GNN architectures on the property graphs. Their bars show the similarities. For the rest of the models, resource  errors (the red bars) manifest the lowest performance among the 5 vulnerability types. One possibility is that resource allocation and free can be located far apart in the code, and the transformer models cannot capture such long range dependencies. Another possibility is that  such errors cover a variety of resources, and the training dataset may not contain sufficient data for each resource for the model to extract the patterns.

The combined model (the brown bar) is generally less performant compared to the models trained with a specific type of vulnerability, but for some vulnerability types like Input validation and Resource errors, the combined model can perform better. For example, for CodeBERT, the combined model achieved higher F1 compared to all of the other 5 models. The circles inside the brown bar show that Privilege escalation and Resource errors reported relatively low accuracy, but for all of the vulnerability types, the combined model reported better performance compared to the type specific models.



Analyzing cross-bugtype performance, we found that for the most of time, cross-bugtype detection reports much lower performance except LineVul, implying that different vulnerability types represent different data distributions.  LineVul seems to handle value errors very well. When applying models trained with other vulnerabilities, value errors reported higher performance than the same-bug performance. 

\newcommand{\bof}{\textsc{Value Propagation}\ }
\newcommand{\inp}{\textsc{Input Validation}\ }
\newcommand{\res}{\textsc{Resource Allocation/Free}\ }
\newcommand{\val}{\textsc{Value Propagation}\ }
\newcommand{\priv}{\textsc{Privilege escalation/Authorization}\ }

\vspace{0.1cm}

\noindent{\bf RQ3} Are programs with certain code features harder to be predicted correctly by the current vulnerability detection models, and if so, what are those code features?

\vspace{0.1cm}
\noindent{\bf Motivation:} Here, we investigated whether we can characterize the programs that cannot be predicted well and are \ben{``hard''} to deep learning models, and whether different models agree on such difficulties. Knowing what programs we cannot handle gives a good target for our future work to improve upon. In program analysis, we know that certain features are hard to handle, such as loops and pointers. We want to know whether these features are also hard for deep learning.



\vspace{0.1cm}
\noindent{\bf Study Setup:} In the first step, we prepared a list of code features for investigation. We think it is interesting to compare with program analysis tools regarding what types of programs are hard to handle. So our approach is to list code features that are important to program analysis, and then check if they also made a difference for deep learning tools.


We obtained a total of 12 code features. Some are control flow related, e.g., the structures of {\it while}, {\it for}, {\it if}, {\it goto}, {\it call} and {\it switch} as well as {\it unconditional jumps} of break, continue, return;  some are data structure and pointer related, e.g., {\it arrays} and {\it pointers} (including the field accesses); and finally some are auxiliary structures such as {\it comment} and {\it macro}. Based on this feature list, we applied the {\it tree-sitter}~\footnote{\url{https://tree-sitter.github.io/tree-sitter/}} parser to count the frequency of code features present in each \ben{function}.

To understand whether certain code features make deep learning models harder to predict, we used a {\it multivariate logistic regression (LR)} model (see Eq. 1) to associate the code features with the likeliness of a \ben{function} being predicted correctly. If a \ben{function} with certain code features is more likely predicted correctly, we consider it as easier to deep learning, and vice versa.  
Given a \ben{function} with a particular feature composition, $Y$ in Eq. 1 reports the predicted probability that the deep learning model will predict it correctly. $x_i$ is the count of each code feature in the \ben{function}. $\beta_i$s are the coefficients learned from the data. Each $\beta_i$ is associated with one code feature $x_i$.
When $\beta_i$ is negative, the value $\beta_i*x_i$ decreases the predicted probability of correctness, so we term these features to be \textit{difficult} for the model. Likewise, code features with positive coefficients are termed to be \textit{easy}. Meanwhile, a large $\beta_i$ implies that the increase in the count of code feature $x_i$ greatly increases the predicted probability of correct prediction, and vice versa.

\begin{equation}
    \label{eq:1}
    Y = \sigma (\sum_i \beta_i * x_i + \beta_0)
\end{equation}

We trained the LR model on the predictions made on the validation set, and then used the trained LR model to find difficult/easy examples from the test set. To quantify the difficulty of an example in test set, we used the logit input to the sigmoid function in the LR model: $\ell(x) = \displaystyle\sum_i \beta_i * x_i$. We denote the negation of this quantity as the \textit{difficulty score}. A \ben{function} with a higher difficulty score is expected to be more likely predicted incorrectly than a \ben{function} with a lower difficulty score. To evaluate the effectiveness of this LR model, we selected the top and bottom 10\% of examples in the test set, sorted by their difficulty scores. We then evaluate the predictions of our LR model by comparing the model performance on the easy and difficult datasets we selected.






It should be noted that initially we have tried statistical significance tests and correlation-based methods to link code features with the model accuracy. Then, we realized that these approaches consider only one feature a time. For example, the \ben{functions} with 3 loops, 400 pointers, and 500 macros perform well, while \ben{those} with 300 loops, 300 pointers and 100 macros perform bad. We cannot conclude whether the performance difference is due to the loops, pointers or macros. On the other hand, the LR model incorporates multiple code features at once and considers the effect of combinations of the features.

\begin{figure}
    \centering
    \includegraphics[width=0.45\textwidth]{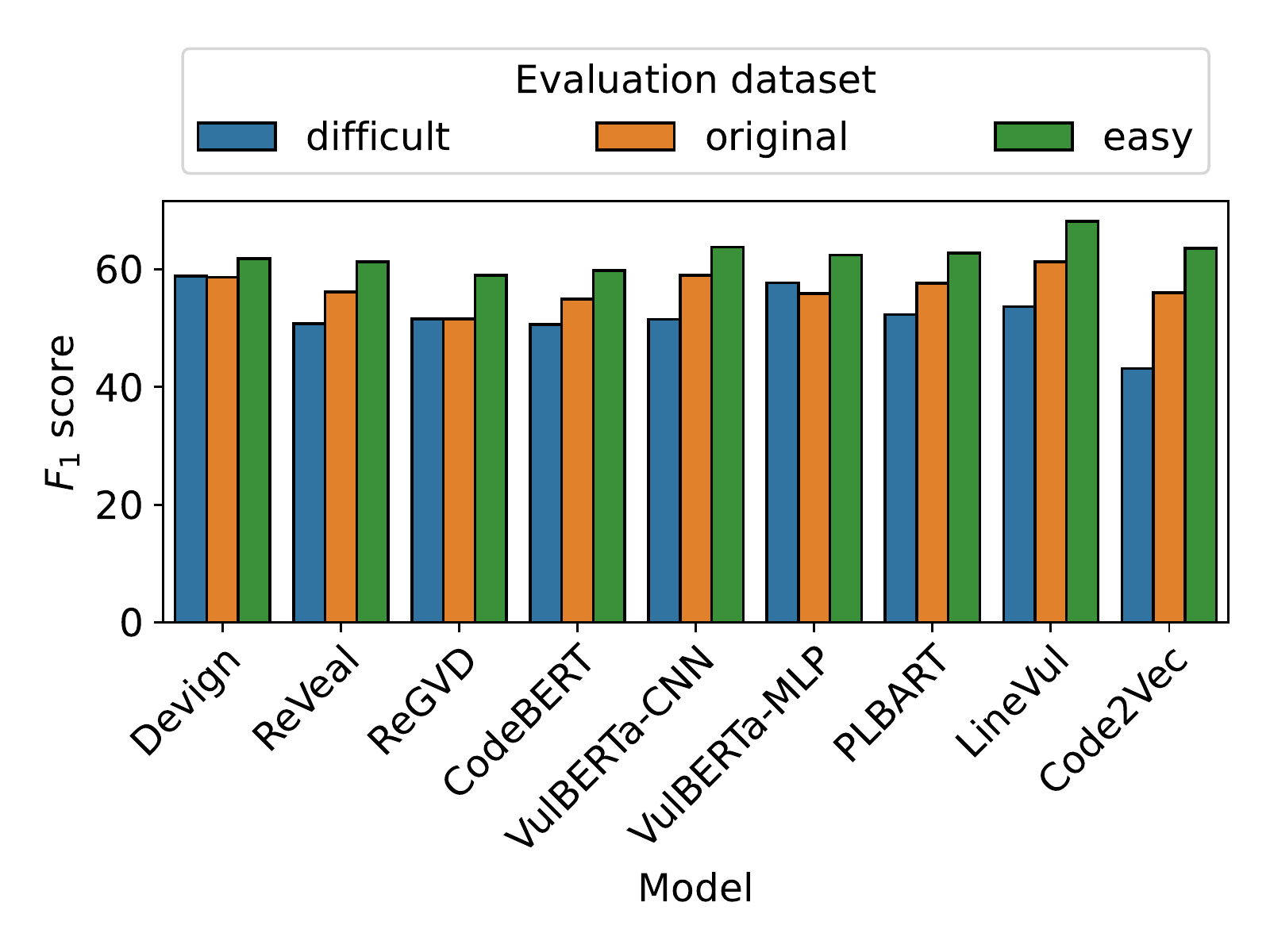}
    \caption{Comparative performance on evaluation sets selected according to LR model difficulty score, averaged over 3 random seeds on the Devign dataset. \ben{``Original''} is the performance on the original test set, reported in Table \ref{tab:reproduction}.}
    \label{fig:rqa1-lr-validation}
\end{figure}

\vspace{0.1cm}
\noindent{\bf Findings:} Figure \ref{fig:rqa1-lr-validation} shows that all 9 models performed better on the easy dataset than on the difficult set. The average difference between easy/difficult performance was 10.3\% for all models. For the majority of models (7 out of 9), the original test set performance lies between the performances of difficult and easy sets. These results demonstrate that the LR model and difficulty score are effective for choosing difficult and easy examples for the deep learning models.

\begin{figure}
    \centering
    \includegraphics[width=0.5\textwidth]{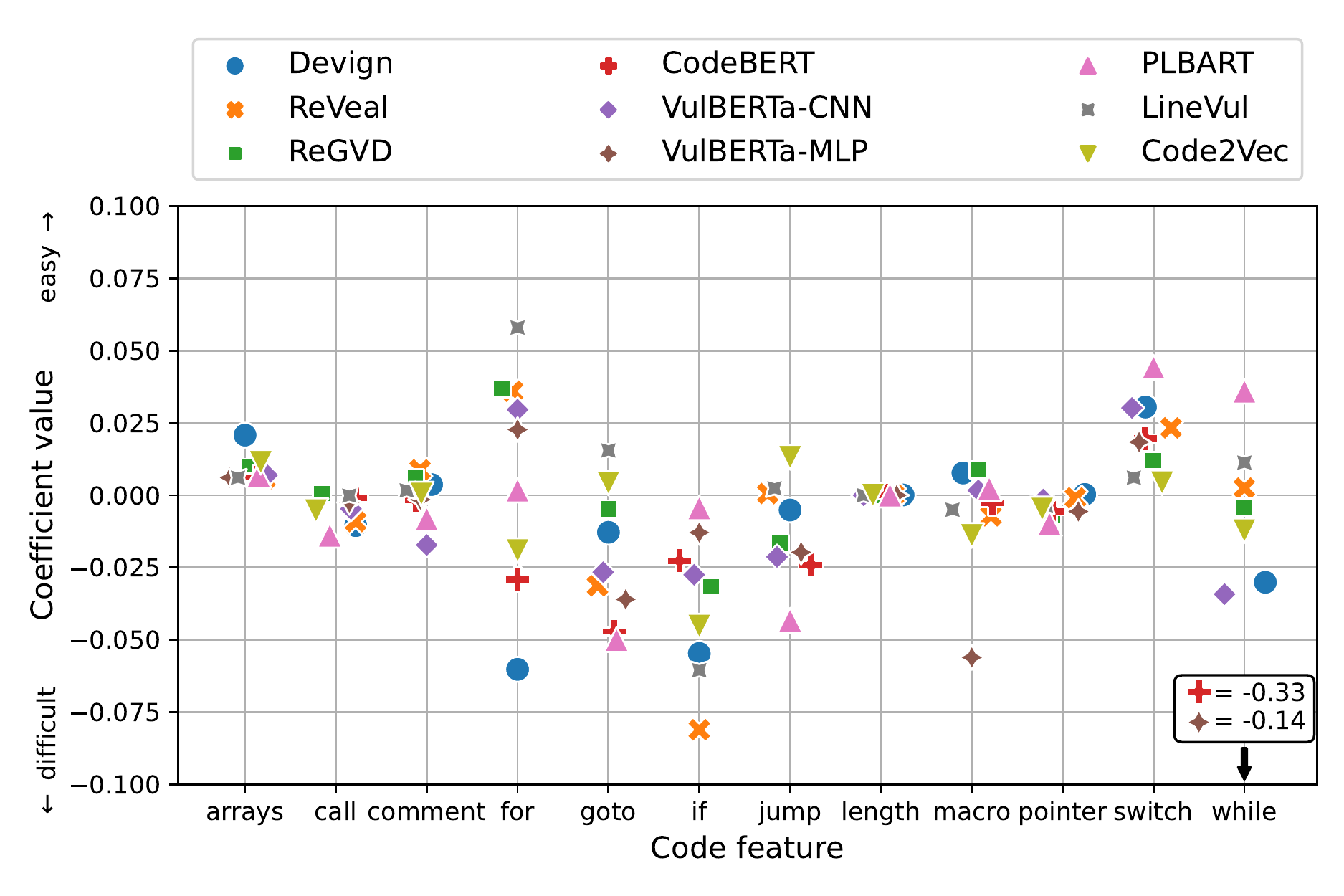}
    \caption{Coefficients of LR models trained on the stable examples from the Devign dataset.}
    \label{fig:rqa1-lr-coefficients}
\end{figure}
     

Figure \ref{fig:rqa1-lr-coefficients} plots the coefficients of each code feature in the LR model trained for each deep learning model.
We found that the dots for the features \verb|call|, \verb|length|, and \verb|pointers| are grouped together for all the models, implying that all the models agreed on the importance of these features. Interestingly, all of these dots are located near 0, which indicates that the features did not have a large effect. On the other hand, the features \verb|for|, \verb|goto|, \verb|if|, \verb|jump|, \verb|switch|, and \verb|while| varied the across models. These are all control flow related structures.


We also observed that for all the models, \verb|arrays| and \verb|switch| are associated with the positive coefficients, and that for the majority of models,  \verb|if|, \verb|goto|, and \verb|while| fall into the negative ranges. In particular, \verb|if| had a negative coefficient for all the models. The high number of {\tt if}, {\tt goto}, and {\tt while} in a program indicates its high cyclomatic complexity~\cite{cyclomatic}, and this type of program is also challenging for program analysis.
Especially, property graph-based models like Devign use control flow information, so it makes sense that features {\tt for}, {\tt goto}, {\tt jump} and {\tt while} were all negative (hard).

\subsection{The Training Data}\label{sec:data}


\noindent{\bf RQ4} Can increasing the dataset size help improve the model performance for vulnerability detection?
	

\vspace{0.1cm}
\noindent{\bf Motivation:} High-quality vulnerability detection data are hard to obtain. In the past, we used automatic labeling methods such as static analysis and mining change commits. These approaches can introduce incorrect labels~\cite{are_we_there_yet}. We also used manual labels~\cite{zhou_devign_2019}, which are slow to produce~\cite{lin_software_2020}. This research question helps us understand whether currently available datasets are large enough to train the models, and whether increasing the dataset size can significantly improve the model performance.

\vspace{0.1cm}
\noindent{\bf Study Setup:} To investigate this RQ, we combined the \devign and \msr datasets, namely,  {\it Imbalanced-dataset}. Since the projects used in \devign overlap with the projects used in MSR, we excluded 82 duplicated examples where the commit IDs are matched. This generates a dataset of 194,285 examples in total. Some of the published models are originally tuned on the balanced model such as Devign, so we also constructed a {\it Balanced-dataset} of 45,363 examples by taking all the vulnerable examples in MSR and then randomly undersampling an equal number of non-vulnerable examples. For each dataset, we held out 10\% data as the test set for all the models. Then we prepared 10\%, 20\% ... 90\%, 100\% of the rest of the data to train 10 models to observe how the F1 score for test set changes when the dataset size increases. In addition, we prepared two small datasets of 1\% and 5\% of the total data to experiment what is the minimum amount of the data needed for the model being able to learning something.



\begin{figure}
    \centering
    \begin{subfigure}[b]{0.45\textwidth}
        \includegraphics[width=\textwidth,keepaspectratio]{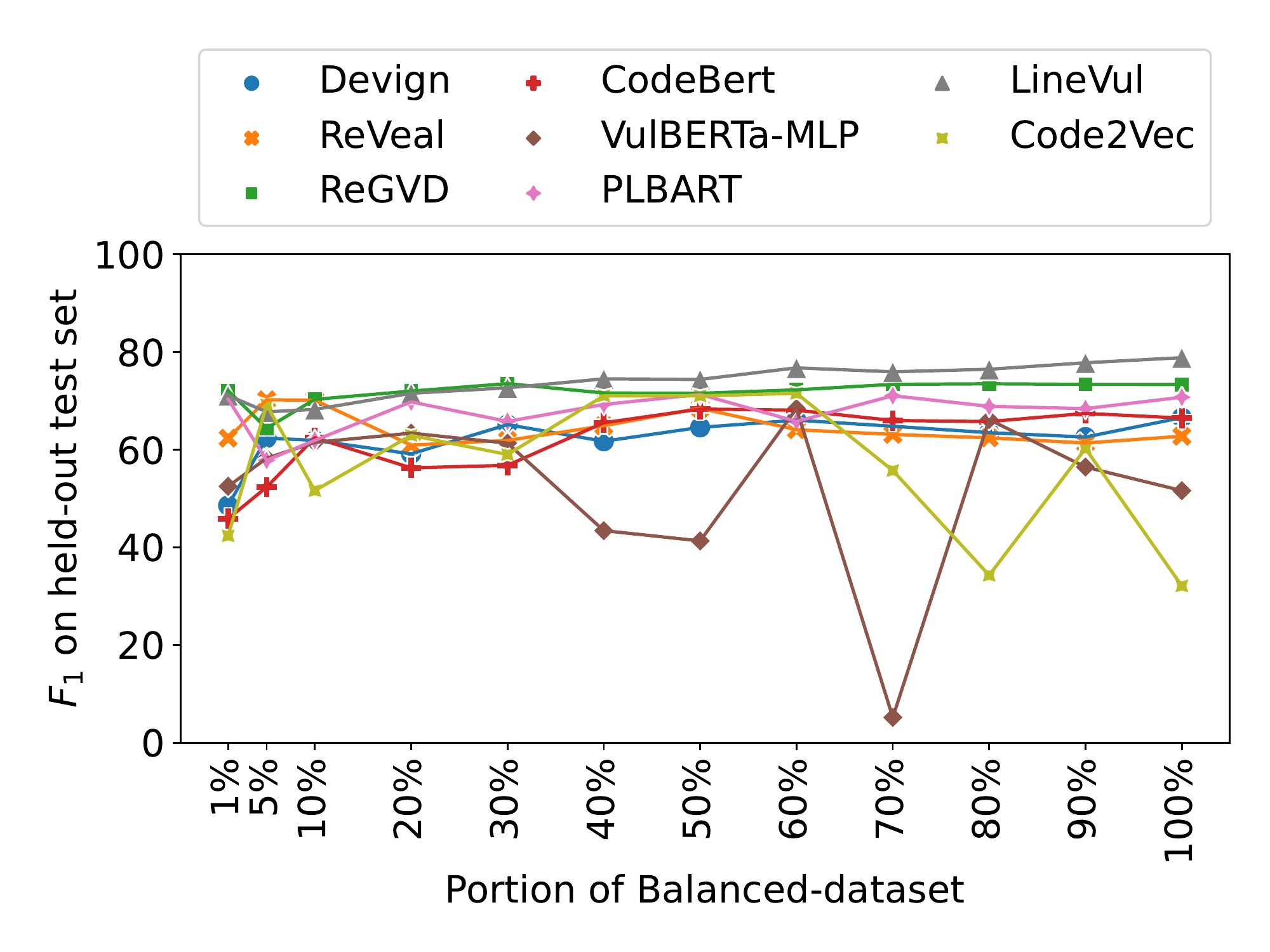}
        \caption{Results on Balanced-dataset. 100\% dataset size = 45,363.}
        \label{fig:rqb1}
    \end{subfigure}
    \hfill
    \begin{subfigure}[b]{0.45\textwidth}
        \includegraphics[width=\textwidth,keepaspectratio]{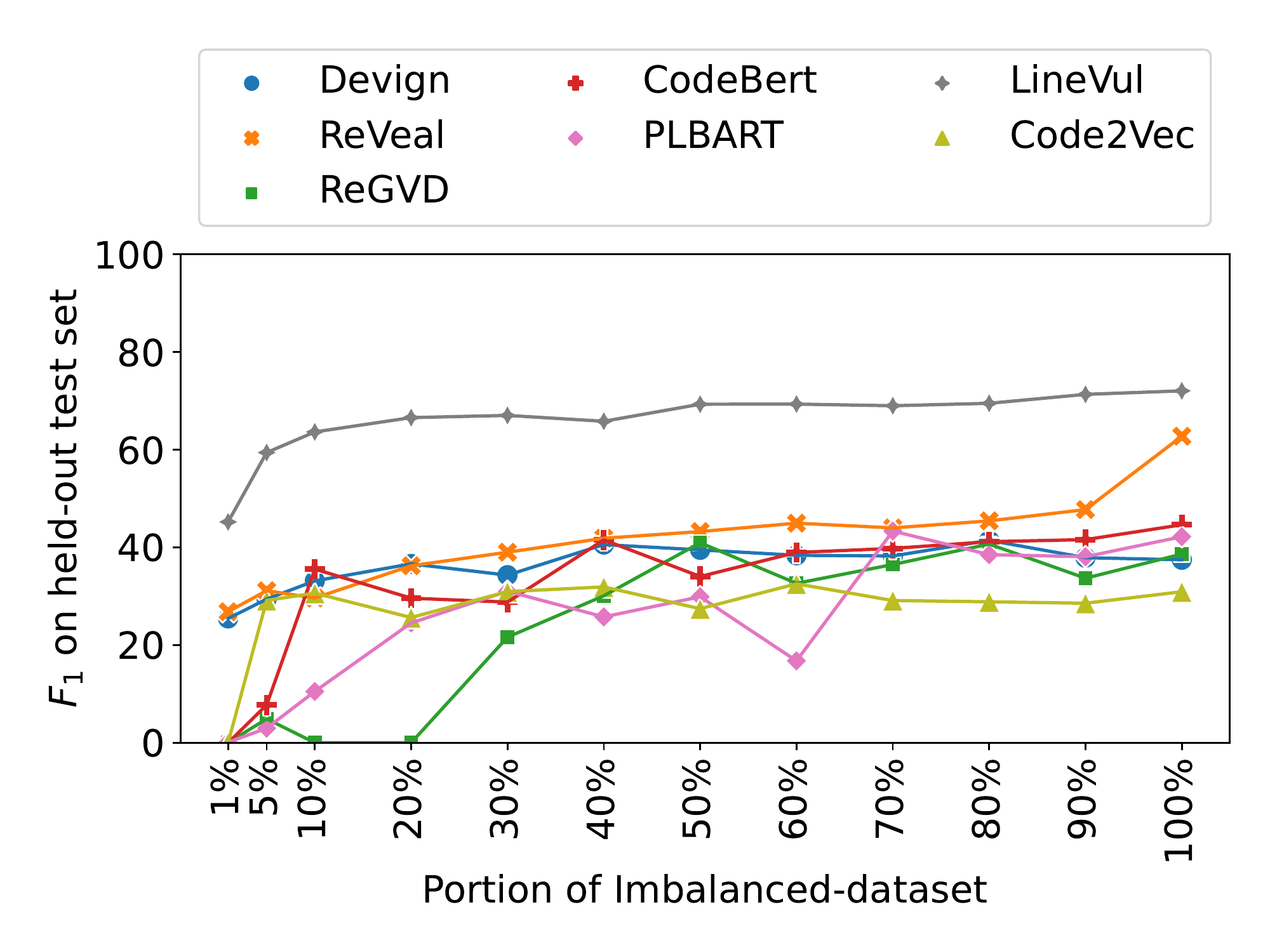}
        \caption{Results on Imbalanced-dataset. 100\% dataset size = 194,285.}
        \label{fig:rqb1-imbalanced}
    \end{subfigure}
    \caption{F1 score on a held-out test set when models are trained with increased portions of the training dataset.}
    \label{fig:rqb1-all}
\end{figure}

\vspace{0.1cm}
\noindent{\bf Findings}: We summarized our results in Figure~\ref{fig:rqb1-all}. Figures~\ref{fig:rqb1} and ~\ref{fig:rqb1-imbalanced} show a similar trend. Generally, all the models increased in performance when we added more data. However, the improvement was not significant. Comparing 100\% data with 10\%, the F1 on test set reported no difference when we take an average over all the models for the Balanced dataset. For the  Imbalanced dataset, the value of F1 score improved 0.16 on average.


In Figure~\ref{fig:rqb1}, among the models, only LineVul showed consistent improvement when we added 10\% more data each time. All other models fluctuated when we increased the training dataset, which indicates that the increasing data does not always bring benefit. For example, for VulBERTa-MLP, the F1 value drops an average 0.1 over the course of increasing the dataset sizes. The performance of the model trained with 100\% of the dataset is 0.08 less than the performance of the model trained with 10\% of the dataset. It seems that other factors rather than the dataset size has played a much significant role in terms of performance. In Figures~\ref{fig:rqb1-imbalanced}, ReVeal is the model that improves the most with the increased dataset. At 100\% dataset, ReVeal is catching up the best model LineVul. However, Devign, which has a similar architecture of using GNN on the property graph, does not show this benefit of additional data.

The experiments on models with the small datasets (consisting of 1\% and 5\% total data) showed that surprisingly, we can bring the models up to good performance only using 5\% (about 2268 data points and 1134 vulnerable examples) for most of the models except CodeBERT, when learning with the balanced data. When learning with the imbalanced data, the turning point comes a little later. For example, ReGVD and CodeBERT require 50\% (96.4 k) and 30\%  (57.8 k) of the total data. The other models need about 5--10\% of the data (9.7--19.4 k) to reach a high point. Interestingly, this dataset has 10.8\% vulnerable examples; that said, the models needed about \ben{1048-2095} vulnerable examples to achieve good performance---a comparable amount of vulnerable examples to the setting of the Balanced dataset.

\vspace{0.1cm}
\noindent{\bf RQ5}  How does the project composition in the training dataset affect the performance of vulnerability detection models?
\vspace{0.1cm}

\noindent{\bf Motivation:} In this RQ, we aim to further understand how to compose a good training dataset for vulnerability detection. Specifically, we are interested to know whether the diversity of the projects in the training dataset helps. We are also interested to learn whether different projects indeed represent different distributions such that when the test and training data come from the same projects, the model can significantly perform better, and when the training and test data are from different projects, whether the models can generalize over unseen projects.



\vspace{0.1cm}
\noindent{\bf Study Setup:} We designed two experiments for this study. In the first experiment, we prepared a {\it non-diverse} training dataset and a {\it diverse} training dataset, and compared the models trained with the two datasets on the same test set. In the MSR dataset, we found that {\it Chrome} contains 76k examples and is the largest among all the 310 projects. We used it as the non-diverse dataset. We performed this experiment in a 5-fold cross validation setting to eliminate the potential biases that may exist when selecting projects. For each fold, we randomly sampled 10 k examples from the MSR dataset as a test set. We then excluded Chrome and the projects used in the test set, and randomly sampled a total of 76 k examples (the same number as Chrome has) from the remaining projects. The average number of projects in the diverse dataset is 50.6 across 5 folds.

In the second experiment, we prepared a {\it mixed-projects} setting where the test set is separated from the training set without considering the source project, and some examples in the training and test sets may originate from the same projects. This is the setting where most of our deep learning papers are evaluated with. We also constructed a {\it cross-projects} setting where the test set examples must originate from different projects than the projects represented the training set. This setting helps us understand whether we will have a significant performance degradation when using an off-the-shelf trained deep learning vulnerability detection models that have not seen the test projects. 

We also used the MSR dataset in a 5-fold cross validation setting. For each fold, we first constructed a test set for the cross-project setting by including all the examples from randomly chosen projects, until the set contained at least 10k examples. Because each project had a different number of examples, the resulting set was slightly larger than 10k examples. We then constructed a test set for the mixed-project setting by randomly partitioning the remaining examples into test (10 k), validation (10 k), and training (the remaining examples, about 158 k) sets.  We trained the model, using the 158 k training examples and 10 k validation set, then ran it on the test sets for both the cross-project setting and the mixed-project setting.

\begin{figure}
    \centering
     \begin{subfigure}{0.45\textwidth}
        \includegraphics[width=\textwidth]{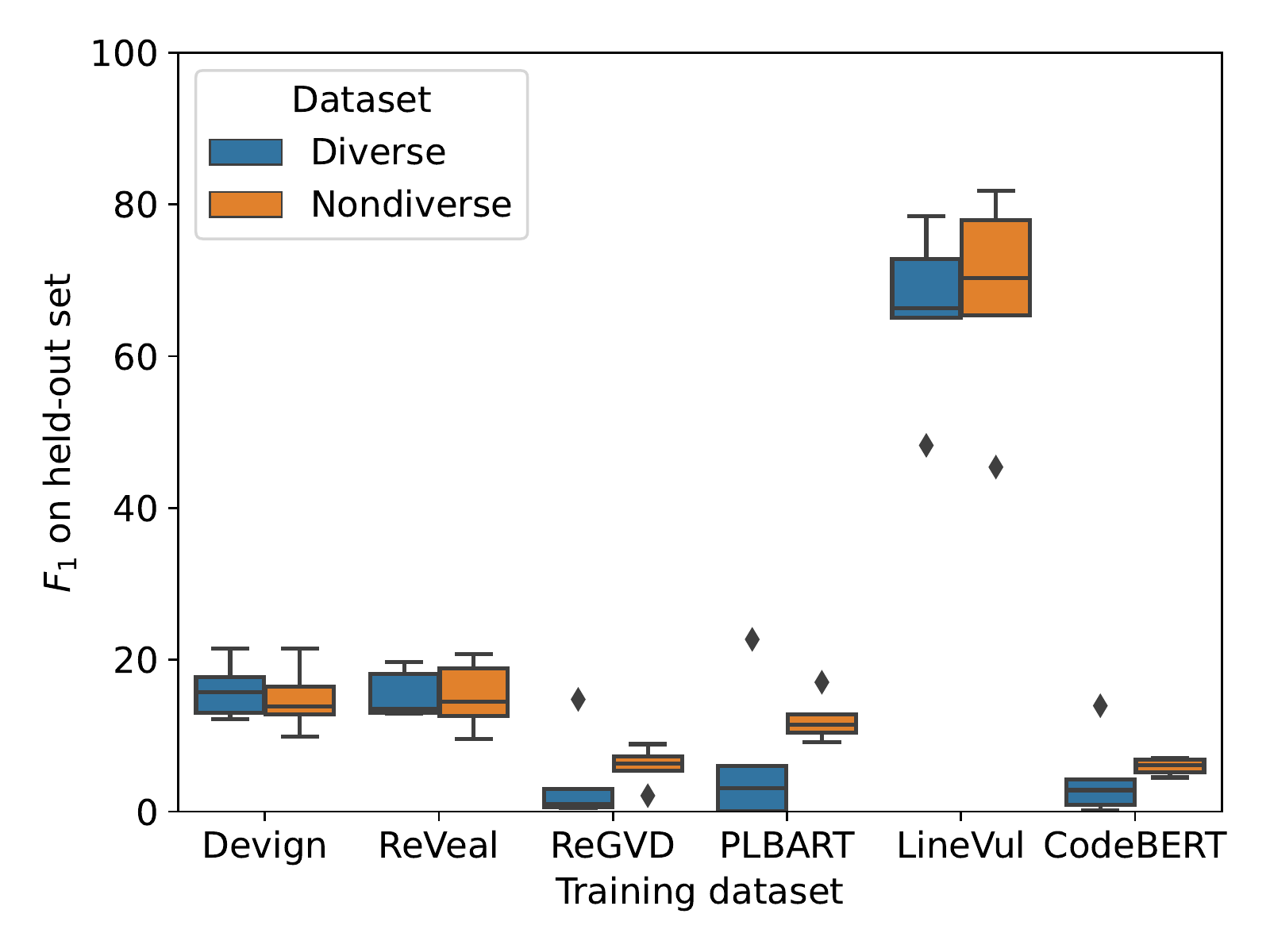}
        \caption{Diverse vs. non-diverse performance}
        \label{fig:rqb3-diversity}
     \end{subfigure}
     
       \begin{subfigure}{0.45\textwidth}
        \includegraphics[width=\textwidth]{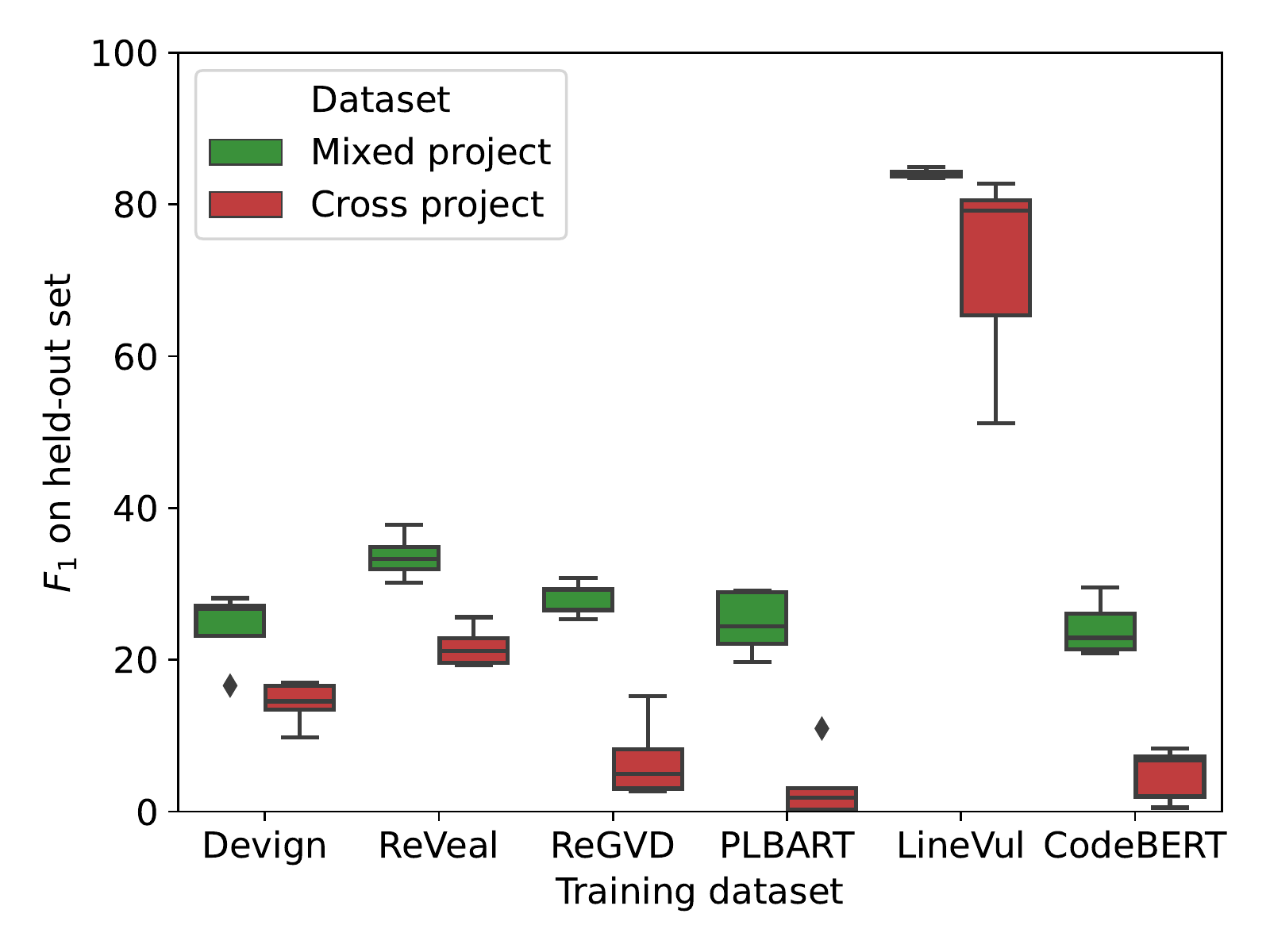}
        \caption{Mixed-project vs. cross-project performance}
        \label{fig:rqb3-crossproject}
     \end{subfigure}
    \caption{Studies on project composition in training data. The bar shows mean F1 and the interval shows standard deviation.}
    \label{fig:rqb3}
\end{figure}

\noindent{\bf Findings:} The results for the first experiment are presented in Figure~\ref{fig:rqb3}. We used the boxplot to summarize the results of 5-fold cross-validation. To our surprise, we found that for all the models, the diverse training set does not provide any benefits compared to the training set that only consists of {\it Chrome}. In fact,  5 out of 6 models reported a higher median performance when trained on non-diverse data than on diverse data.

For the second experiment, Figure \ref{fig:rqb3-crossproject} shows that mixed projects perform significantly better than cross projects for all the models (the average and the largest differences in F1 score were 0.11 and 0.32 respectively). This implies that seeing the data from one project can indeed help predict other data from the same project. Vulnerability detection can greatly benefit from customized trained models compared to directly used already trained off-the-shelf models. For LineVul, the 5 folds reported very different performances in the cross-project setting, indicating that given a target test set, some projects are more useful for training than others. The results also imply that the models may learn to detect vulnerabilities from project-specific attributes of the dataset, such as style, language features or naming conventions. We believe that this motivates further research into causal detection of bugs for generalization.



\mycomment{
\vspace{0.2cm}
\noindent \fbox{
	\parbox{0.95\linewidth}{ {\bf Summary}: For most models, the test F1 score \ben{was} lower when applied to unseen projects compared to when the model used same projects for training and testing. Increasing the project diversity in training datasets \ben{was} not helpful for the cross-project generalization.
}
}
}
\vspace{0.1cm}

\subsection{Internals of Deep Learning}\label{sec:model-interpretation}
\noindent{\bf RQ6} What code information do the models use for prediction?  Do the models agree on the important features?

\vspace{0.1cm}
\noindent{\bf Motivation:} Recent deep learning vulnerability detectors have achieved high performance, e.g., LineVul reported 91\% F1 score. We want to know why these tools can perform well, and whether the model has used any aspects of semantics of the vulnerability to make decisions. For example, to detect buffer overflows, a semantic-based program analysis tool identifies the dependent statements and reasons about the string lengths and buffer sizes. We also want to investigate whether different models agree on what are the important features. 



\vspace{0.1cm}
\noindent{\bf Study Setup:}  We surveyed a set of SOTA deep learning interpretation tools, especially for GNN and transformer architectures. We used GNNExplainer\cite{GNNExplainer} for Devign and ReGVD, and LIT\cite{LIT_2020,Integrated_Gradients_2017,LIME_2016} for LineVul, VulBERTa-CNN, VulBERTa-MLP, CodeBert and PLBART, as among all the tools we investigated, these two tools can work with the most for our models. In our data reproduction package, we documented the reasons why the rest of the models cannot work with  GNNExplainer and LIT, as well as the other intepretability tools we have tried.

To explain the models, both of GNNExplainer and LIT provide scores to measure the importance of the code features. GNNExplainer gives a score for each edge in the graph, and LIT gives a score for each token in the program. To compare the results reported by GNNExplainer and LIT, we performed the following {\it normalization} on the output of the tools. For GNNExplainer, we calculated the score of each node by taking the average scores of all of its incident edges, as what has been done in~\cite{ivdetect}. Each token in the node will use this score. For both GNNExplainer and LIT,  we calculated the score for each source code line by summing up the scores for all the tokens of that line, following the literature~\cite{linevul}. For each example in the test dataset, we selected the top 10 highest scored lines, as done in ~\cite{ivdetect,linevul}, to form the {\it important feature set}, denoted as {\it I}. We considered these 10 lines as the most important code features the model used to make a decision. 

To measure the similarity of the two important feature sets $I_A$ and $I_B$ reported by the models A and B, we calculated the intersection $I_{AB}=I_A\cap I_B$. We also used {\it Jaccard index}\cite{jaccard} as another metric, defined as:

\begin{equation}
    \label{eq:1}
    J(I_A,I_B) = \frac{|I_A \cap I_B|}{|I_A \cup I_B|}
\end{equation} 
To report the similarity, we first compute $I_{AB}$ and $J(I_A,I_B)$ for each program in the test set, and then we took the average for $I_{AB}$ and $J(I_A,I_B)$ respectively.

We sampled examples from the following groups for manual inspection: 1) the example is vulnerable, and the model correctly detected it (correct);  2) the example is non-vulnerable, and the model predicted as vulnerable (false positive); and 3) the example is vulnerable, and the model predicted as non-vulnerable (false negative). 


\vspace{0.1cm}
\noindent{\bf Findings:} In  Table~\ref{tab:interpretability_line}, we reported the similarity of the important feature sets from every model pair. Our results show that Linevul and ReGVD have a maximum overlap among all the model pairs. Among the 10 lines ranked as the important features, the two models shared an average of 6.88 lines.  We found it interesting that although the models can have a lot of disagreement on individual predictions (See Table IV), the code information they used overlap. All the model pairs have at least 3 lines in common for the important features. Devign, as the only GNN model based on the property graph, has a low overlap with the other models and the lowest is with PLBART, on average 3.38 lines. PLBART used a different transformer architecture compared to the other transformer model, and also reported low overlaps with other transformer models. 

We have also analyzed the corrected predicted examples using the same approach as Table~\ref{tab:interpretability_line}. We found that the important feature sets have more overlaps for the corrected examples, e.g., Linevul and ReGVD still have the most in common, sharing 7.29 lines in the important feature sets.

\begin{table}[htbp]
    \centering
    \caption{The similarity of important feature sets between every two models measured by $I_{AB}$ (reported in blue) and $J(I_A,I_B)$ (in black). The max and min values are bold.}
    \resizebox{\columnwidth}{!}{\begin{tabular}{|l|c|c|c|c|c|c|c|}
        \hline
         Model  &  L-Vul & CodeB & PLBART & Devign &ReGVD & V-CNN & V-MLP\\ \hline\hline
        L-Vul & - & { 0.58}&0.31 &0.38& \bf 0.60& 0.32 &0.40\\\hline
      CodeB & {\color{blue} 6.84} & - &0.37&0.35& 0.50& 0.31 &0.40\\\hline
        PLBART &  {\color{blue} 4.08} &  {\color{blue} 4.89} & - & {\bf 0.24}& 0.27 & 0.30 & 0.28\\\hline
        Devign &  {\color{blue} 4.90} &  {\color{blue} 4.54} & {\color{blue}\bf 3.38}& - &0.42 & 0.27 & 0.30\\\hline
        ReGVD &  {\color{blue} \bf 6.88} & {\color{blue} 5.86}& {\color{blue} 3.58}& {\color{blue} 5.36}& - & 0.33 &0.40\\\hline
        V-CNN &  {\color{blue} 4.20} & {\color{blue} 4.06}& {\color{blue} 3.88}& {\color{blue} 3.87}&\color{blue} 4.35 & - &0.30\\\hline
        V-MLP &  {\color{blue}  4.83} & {\color{blue} 4.74}& {\color{blue} 3.71}& {\color{blue} 3.95}&\color{blue}  4.83 & \color{blue} 3.97 &-\\\hline
        
    \end{tabular}}
    \label{tab:interpretability_line}
\end{table}

\begin{table}[]
\caption{The frequently highlighted code features}\label{feature}
\resizebox{\columnwidth}{!}{
\begin{tabular}{|l||c|c|c|c|c|c|c|c|}
\hline
Model      & \multicolumn{1}{c|}{error} & \multicolumn{1}{c|}{print} & \multicolumn{1}{c|}{alloc} & \multicolumn{1}{c|}{for} & \multicolumn{1}{c|}{memset} & \multicolumn{1}{c|}{memcpy} & \multicolumn{1}{c|}{while} & \multicolumn{1}{c|}{if} \\ \hline\hline
L-Vul      & .035                       & .021                       & .017                       & .042                     & .003                        & .003                        & .005                       & .115                    \\ \hline
CodeB      & .027                       & .015                       & .015                       & .032                     & .004                        & .002                        & .004                       & .111                    \\ \hline
PLBART     & .025                       & .012                       & .011                       & .034                     & .003                        & .001                        & .005                       & .114                    \\ \hline
Devign     & .023                       & .012                       & .014                       & .069                     & .002                        & .002                        & .007                       & .147                    \\ \hline
ReGVD      & .034                       & .027                       & .022                       & .038                     & .004                        & .005                        & .005                       & .132                    \\ \hline
V-CNN      & .024                       & .014                       & .011                       & .028                     & .002                        & .002                        & .004                       & .116                    \\ \hline
V-MLP      & .031                       & .018                       & .015                       & .020                     & .003                        & .003                        & .004                       & .095                    \\ \hline\hline
Func       & .010                       & .010                       & .010                       & .026                     & .002                        & .002                        & .004                       & .108                    \\ \hline\hline
I/F & 2.79                       & 1.78                       & 1.52                       & 1.47                     & 1.39                        & 1.21                        & 1.12                       & 1.10                    \\ \hline
\end{tabular}
}
\end{table}

From our manual inspection, we observed that the models commonly highlighted code lines of {\tt for}, {\tt if}, and {\tt while}, as well as the function signatures as important features. The models also often \ben{highlighted} memory operations of {\tt alloc}, {\tt memset}, and {\tt memcpy}, as well as the lines which print error messages containing \ben{{\tt error} or {\tt printf}}. To confirm this observation, we performed profiling on the code using these keywords \ben{and reported the results} in Table~\ref{feature}. Using {\tt error} as an example, without loss of generality, \ben{the first 7 rows report the probability of {\tt error} occurring in the important feature set for each model} (total number of  {\tt error} in the important feature set/total number of lines in the important feature set).  In Row {\it Func}, we show the probability of {\tt  error} occurring in a function (total number of {\tt error}/total number of lines in a function). Comparing the two, we show that the probability of {\tt error} occurring in the important feature sets is 2.79 times of the probability of {\tt error} occurring in the program on average, shown in Row {\it I/F}. This implies that {\tt error} is preferred to be selected into the important feature sets. Among all the features,  {\tt error}, {\tt print}, and {\tt alloc} ranked the highest ratios.


    
Our second observation is that the transformer models sometimes made predictions without seeing the root cause. This is because the transformer models take a fixed-size input, and some code, sometimes including the root cause, is truncated. Interestingly, those models are still able to correctly predict whether a function is vulnerable with high F1 score.

Third, we inspected the vulnerabilities that all the models missed and studied the important feature sets used to detect such vulnerabilities. We found that these vulnerabilities are very application specific. The bugs are missed may because there are not sufficient training data for such types of bugs. 

\newcommand{\highlight}[1]{#1} 
\newcommand{\diffadd}[1]{\textcolor{green!60!black}{#1}}
\newcommand{\diffremove}[1]{\textcolor{red}{#1}}
\def\mystrut(#1,#2){\vrule height #1pt depth #2pt width 0pt} 

\begin{listing}
    \begin{minted}[escapeinside=@@,breaklines,fontsize=\footnotesize,breakanywhere,linenos,highlightcolor=yellow!50,highlightlines={1,5,6,9,13,16,17,18,19,21},numbersep=3pt]{C}
static int asf_read_ext_content(AVFormatContext *s, const GUIDParseTable *g) // (3)
{
    ASFContext *asf  = s->priv_data;
    AVIOContext *pb  = s->pb;
    @\highlight{uint64\_t size    = avio\_rl64(pb);}@ //(7)
    @\highlight{uint16\_t nb\_desc = avio\_rl16(pb);}@ //(8)
    int i, ret;
    for (i = 0; i < nb_desc; i++) {
        @\highlight{uint16\_t name\_len, type, val\_len;}@ //(5)
        uint8_t *name = NULL;
        name_len = avio_rl16(pb);
        if (!name_len)
            @\highlight{return AVERROR\_INVALIDDATA;}@ //(6)
        name = av_malloc(name_len);
        if (!name)
            @\highlight{return AVERROR(ENOMEM);}@ //(9)
        @\highlight{avio\_get\_str16le(pb, name\_len, name,}@ //(4)
                         @\highlight{name\_len);}@ //(2)
        @\highlight{type    = avio\_rl16(pb);}@ //(10)
        val_len = avio_rl16(pb);
        @\diffremove{\highlight{if ((ret = process\_metadata(s, name, name\_len, val\_len, type, \&s->metadata)) < 0)}}@ //(1)
        @\diffadd{ret = process\_metadata(s, name, name\_len, val\_len, type, \&s->metadata);}@
        @\diffadd{av\_freep(&name);}@
        @\diffadd{if (ret < 0)}@
            return ret;
    } 
}

    \end{minted}
    \caption{Memory leak successfully detected by LineVul. The top-10 lines reported by LIT are highlighted in yellow.}
    \label{lst:linevul-memoryleak-success}
\end{listing}

In Listing~\ref{lst:linevul-memoryleak-success}, we show an example where the prediction is correct, but the features used are not causal. 
The example contains a memory leak vulnerability, and LineVul predicted the example as vulnerable. The memory allocated to {\tt name} at line~14 is never released. The patch at line~23 showed a fix. The important feature set (top-10 lines) reported by LIT are highlighted in yellow. We can see that it includes the \ben{``patterns''} we have discussed, including function signature at line 1, the lines that contain \ben{{\tt ERROR}}, e.g., lines 13 and 16,
as well as the {\tt if} statement at line 21. We also see that for this project, variable {\tt name\_len} is important and included multiple times. However, none of the 10 lines cover the memory allocation at line~14, which is important to understand this bug.

This example indicates that the models try to capture patterns of a vulnerability, instead of reasoning about the values, and have difficulty capturing long range semantic dependencies in the code. But we also observed in other examples that sometimes, the control structures and memory statements highlighted as important (see Table VII)  can be a part of the dependent statements of the vulnerability, and thus they are useful for inspecting the root cause of the bugs. 


\begin{listing}
    \begin{minted}[escapeinside=@@,breaklines,fontsize=\footnotesize,linenos,breakanywhere,highlightcolor=yellow!50,highlightlines={2,5,6,7,9,10,14,15,16,17},numbersep=3pt]{C}
static int decode_frame(AVCodecContext *avctx,
                        @\highlight{void *data, int *got\_frame, AVPacket *avpkt)}@ // (1)
{
    //...10 lines
    @\highlight{bytestream2\_init(\&s->gb, avpkt->data, avpkt->size);}@ // (5)
    @\highlight{if ((ret = ff\_tdecode\_header(\&s->gb, \&le, \&off))) \{}@ // (6)
        av_log(avctx, AV_LOG_ERROR, "Invalid TIFF header\n"); // (3)
        return ret;
    @\highlight{\} else if (off >= UINT\_MAX - 14 || avpkt->size < off + 14) \{}@ // (4)
        av_log(avctx, AV_LOG_ERROR, "IFD offset is greater than image size\n"); // (2)
        return AVERROR_INVALIDDATA;
    }
    s->le          = le;
 // @\highlight{TIFF\_BPP is not a required tag and defaults to 1}@ // (10)
    @\highlight{s->bppcount    = s->bpp = 1;}@ // (9)
    @\highlight{s->photometric = TIFF\_PHOTOMETRIC\_NONE;}@ // (7)
    @\highlight{s->compr       = TIFF\_RAW;}@ // (8)
    // ...140 lines
    \end{minted}
    \caption{Non-vulnerable code is predicted as vulnerable because of spurious features.
    }
    \label{lst:linevul-spurious}
\end{listing}

Listing \ref{lst:linevul-spurious} shows an example of a non-vulnerable function which LineVul erroneously predicted as vulnerable. The model highlighted the function signature (line 2), lines with ``ERROR'' (lines 7 and 10), initialization routines (line 5), \texttt{if} (lines~6 and 9), and field assignments (lines 15-17), and predicted the function as vulnerable. It showed that making decisions based on the patterns of these structures can lead to mistakes.

\section{Threats to Validity}\label{sec:threats}
Our observations are drawn from the models and data sets available and may not be generalized for deep learning vulnerability detection in general.
\ben{We used both a balanced dataset (Devign) and an imbalanced dataset (MSR) to mitigate this threat. The two datasets both included real-world bugs. Devign is used by most of the models in their evaluation, so we need it to reproduce the models (see Table~\ref{survey}).  However, our datasets may still not be representative of the real-world vulnerability distribution.  We included all the models that we could find and reproduce.}



The grouping in RQ2 is subject to bias in that different researchers may divide vulnerability types differently. Here, two of the authors who have the domain knowledge inspected the CWE list individually and discussed and agreed on the grouping. To mitigate the bias that may be brought in by the specific project compositions, RQ5 performed 5-fold cross-validation. For RQ6, we selected the SOTA model interpretation tools; however, such techniques may not be perfect to identify the important features that the models use. The experiments for RQ2, RQ4 and RQ5 require the models to work with our customized data that are not shipped with the models. We tried different random seeds for any suspicious data we have observed, e.g., when a model reported all 0s or 1s. We excluded such models in our results when tuning could not resolve the issues.








\section{Related Work}\label{sec:related}
Several works have done empirical studies of machine learning based vulnerability detection models. Chakaborthy et al. \cite{are_we_there_yet} studied 4 DL models, and investigated the issues of synthetic datasets, data duplication and data imbalance, and pointed out the use of spurious features, then used these to improve their model design.
Tang et al. \cite{tang_comparative_2020} aim to determine which neural network architectures, vector representation methods, symbolization methods are the best. They surveyed 2 models. Mazuera-Rozo et al.~\cite{mazuera-rozo_shallow_2021} evaluated 1 shallow and 2 deep models on binary classification and bug type (non-binary) classification. 
After we completed our study, we found two related empirical studies.
Lin et al. \cite{lin_deep_2021} evaluated 6 DL models' generalization for 9 software projects. Ban et al. \cite{ban_performance_2019} evaluated 6 machine learning models (1 of which is a neural network) in a cross-project setting with 3 software projects, and also studied training on 2 bug types vs. a single bug type.

Recently, many vulnerability detection models are proposed with a variety of architectures, such as MLP \cite{code2vec_vuln}, RNN \cite{vuldeepecker,sysevr,vuldeelocator,astnn}, CNN \cite{draper,vulcnn}, Transformer \cite{codebert,cotext,syncobert,disco,linevul,plbart,ding2022velvet,vulberta}, and GNN \cite{deepwukong,hoppity,bgnn4vd,ivdetect,hgvul,mvd,acgdp,linevd,zhou_devign_2019,are_we_there_yet,regvd,hellendoorn_global_2020}. For example, Devign used gated graph neural network on property graphs~\cite{Joern}.  LineVul \cite{linevul} used a transformer model pretrained over a large body of diverse open-source projects. ReVeal \cite{are_we_there_yet} applied SMOTE to address the data imbalance issue and triplet loss to learn to maximally separate vulnerable and non-vulnerable code. 

In these papers, most models were evaluated on in-distribution data, where the training set can contain projects and bug types which overlap with the test set.
Russell et al. \cite{draper}, Li et al. \cite{vuldeepecker}, and Xu et al. \cite{acgdp} trained their models to detect specific kinds of vulnerabilities, and all found that some vulnerabilities were more difficult than others.
Hin et al. \cite{linevd} evaluated their model in a cross-project setting by holding out one project at a time and found that the performance was slightly degraded.  Most model evaluations compared different baselines on metrics such as F1, but did not quantify the agreement on the predictions. To the best of our knowledge, our work is the first attempt to characterize the programs and code features which the model cannot predict well.

\section{Conclusions and Future Work}\label{sec:conclusion}







To understand deep learning vulnerability detection models, we performed an empirical study with  6 research questions. We experimentally show that on average, 34.9\% test data have different predictions between runs, and only 7\% of predictions are agreed across 9 models. Vulnerability detection based on a specific type generally performs better than a model built for all vulnerabilities. The model performance does not increase significantly with an increased dataset, and for both balanced and imbalanced datasets, the models start performing well using around 1k vulnerable examples. We developed a logistic regression model that can find programs that are difficult for the model to predict correctly. The explanation tools showed that the models used common features to make predictions, ranging from 3.38-6.88 lines in common per top 10 important lines. We report the code patterns that the models frequently highlighted as important features. In the future work, we plan to further investigate these patterns.

\section{\ben{Acknowledgements}}

\ben{We thank the anonymous reviewers for their valuable feedback. We thank Hongyang Gao for providing computing resources for our experiments. We thank Qi Li for discussing an experiment metric. This research is partially supported by the U.S.
National Science Foundation (NSF) under Award~\#1816352.}

\bibliographystyle{plain} 
\bibliography{main.bib,other.bib}

\begin{thebibliography}{10}

\bibitem{Joern}
Joern.
\newblock ~\url{https://github.com/octopus-platform/joern}.

\bibitem{plbart}
Wasi~Uddin Ahmad, Saikat Chakraborty, Baishakhi Ray, and Kai-Wei Chang.
\newblock Unified pre-training for program understanding and generation, 2021.

\bibitem{code2vec}
Uri Alon, Meital Zilberstein, Omer Levy, and Eran Yahav.
\newblock Code2vec: Learning distributed representations of code.
\newblock {\em Proc. ACM Program. Lang.}, 3(POPL), January 2019.

\bibitem{ban_performance_2019}
Xinbo Ban, Shigang Liu, Chao Chen, and Caslon Chua.
\newblock A performance evaluation of deep‐learnt features for software
  vulnerability detection.
\newblock {\em Concurrency and Computation: Practice and Experience}, 31(19),
  October 2019.

\bibitem{bgnn4vd}
Sicong Cao, Xiaobing Sun, Lili Bo, Ying Wei, and Bin Li.
\newblock {BGNN4VD}: {Constructing} {Bidirectional} {Graph} {Neural}-{Network}
  for {Vulnerability} {Detection}.
\newblock {\em Information and Software Technology}, 136:106576, August 2021.

\bibitem{mvd}
Sicong Cao, Xiaobing Sun, Lili Bo, Rongxin Wu, Bin Li, and Chuanqi Tao.
\newblock Mvd: Memory-related vulnerability detection based on flow-sensitive
  graph neural networks.
\newblock 2022.

\bibitem{are_we_there_yet}
Saikat Chakraborty, Rahul Krishna, Yangruibo Ding, and Baishakhi Ray.
\newblock Deep learning based vulnerability detection: Are we there yet?
\newblock {\em IEEE Transactions on Software Engineering}, 48(9):3280--3296,
  2022.

\bibitem{deepwukong}
Xiao Cheng, Haoyu Wang, Jiayi Hua, Guoai Xu, and Yulei Sui.
\newblock {DeepWukong}: {Statically} {Detecting} {Software} {Vulnerabilities}
  {Using} {Deep} {Graph} {Neural} {Network}.
\newblock {\em ACM Transactions on Software Engineering and Methodology},
  30(3):38:1--38:33, April 2021.

\bibitem{code2vec_vuln}
David Coimbra, Sofia Reis, Rui Abreu, Corina Păsăreanu, and Hakan Erdogmus.
\newblock On using distributed representations of source code for the detection
  of {C} security vulnerabilities, 2021.

\bibitem{hoppity}
Elizabeth Dinella, Hanjun Dai, Ziyang Li, Mayur Naik, Le~Song, and Ke~Wang.
\newblock Hoppity: Learning graph transformations to detect and fix bugs in
  programs.
\newblock In {\em International Conference on Learning Representations}, 2020.

\bibitem{disco}
Yangruibo Ding, Luca Buratti, Saurabh Pujar, Alessandro Morari, Baishakhi Ray,
  and Saikat Chakraborty.
\newblock Towards learning (dis)-similarity of source code from program
  contrasts, 2021.

\bibitem{ding2022velvet}
Yangruibo Ding, Sahil Suneja, Yunhui Zheng, Jim Laredo, Alessandro Morari, Gail
  Kaiser, and Baishakhi Ray.
\newblock Velvet: a novel ensemble learning approach to automatically locate
  vulnerable statements, 2022.

\bibitem{bigvul}
Jiahao Fan, Yi~Li, Shaohua Wang, and Tien~N. Nguyen.
\newblock A {C/C++} code vulnerability dataset with code changes and cve
  summaries.
\newblock In {\em Proceedings of the 17th International Conference on Mining
  Software Repositories}, MSR '20, page 508–512, New York, NY, USA, 2020.
  Association for Computing Machinery.

\bibitem{codebert}
Zhangyin Feng, Daya Guo, Duyu Tang, Nan Duan, Xiaocheng Feng, Ming Gong, Linjun
  Shou, Bing Qin, Ting Liu, Daxin Jiang, and Ming Zhou.
\newblock {CodeBERT}: {A} pre-trained model for programming and natural
  languages.
\newblock {\em CoRR}, abs/2002.08155, 2020.

\bibitem{linevul}
Michael Fu and Chakkrit Tantithamthavorn.
\newblock {LineVul}: A transformer-based line-level vulnerability prediction.
\newblock In {\em 2022 IEEE/ACM 19th International Conference on Mining
  Software Repositories (MSR)}, pages 608--620, 2022.

\bibitem{vulberta}
Hazim Hanif and Sergio Maffeis.
\newblock {VulBERTa}: Simplified source code pre-training for vulnerability
  detection.
\newblock In {\em 2022 International Joint Conference on Neural Networks
  (IJCNN)}, pages 1--8, 2022.

\bibitem{hellendoorn_global_2020}
Vincent~J. Hellendoorn, Charles Sutton, Rishabh Singh, Petros Maniatis, and
  David Bieber.
\newblock Global relational models of source code.
\newblock In {\em International Conference on Learning Representations}, 2020.

\bibitem{linevd}
David Hin, Andrey Kan, Huaming Chen, and M.~Ali Babar.
\newblock {LineVD}: Statement-level vulnerability detection using graph neural
  networks, 2022.

\bibitem{codenet}
IBM.
\newblock Project codenet, 2021.

\bibitem{jaccard}
Paul Jaccard.
\newblock The distribution of the flora in the alpine zone.1.
\newblock {\em New Phytologist}, 11(2):37--50, 1912.

\bibitem{ivdetect}
Yi~Li, Shaohua Wang, and Tien~N. Nguyen.
\newblock Vulnerability detection with fine-grained interpretations.
\newblock In {\em Proceedings of the 29th ACM Joint Meeting on European
  Software Engineering Conference and Symposium on the Foundations of Software
  Engineering}, ESEC/FSE 2021, page 292–303, New York, NY, USA, 2021.
  Association for Computing Machinery.

\bibitem{vuldeelocator}
Zhen Li, Deqing Zou, Shouhuai Xu, Zhaoxuan Chen, Yawei Zhu, and Hai Jin.
\newblock {VulDeeLocator}: A deep learning-based fine-grained vulnerability
  detector.
\newblock {\em {IEEE} Transactions on Dependable and Secure Computing},
  19(4):2821--2837, jul 2022.

\bibitem{sysevr}
Zhen Li, Deqing Zou, Shouhuai Xu, Hai Jin, Yawei Zhu, and Zhaoxuan Chen.
\newblock {SySeVR}: A framework for using deep learning to detect software
  vulnerabilities.
\newblock {\em {IEEE} Transactions on Dependable and Secure Computing},
  19(4):2244--2258, jul 2022.

\bibitem{vuldeepecker}
Zhen Li, Deqing Zou, Shouhuai Xu, Xinyu Ou, Hai Jin, Sujuan Wang, Zhijun Deng,
  and Yuyi Zhong.
\newblock {VulDeePecker}: A deep learning-based system for vulnerability
  detection.
\newblock In {\em Proceedings 2018 Network and Distributed System Security
  Symposium}. Internet Society, 2018.

\bibitem{lin_software_2020}
Guanjun Lin, Sheng Wen, Qing-Long Han, Jun Zhang, and Yang Xiang.
\newblock Software {Vulnerability} {Detection} {Using} {Deep} {Neural}
  {Networks}: {A} {Survey}.
\newblock {\em Proceedings of the IEEE}, 108(10):1825--1848, October 2020.
\newblock Conference Name: Proceedings of the IEEE.

\bibitem{lin_deep_2021}
Guanjun Lin, Wei Xiao, Leo~Yu Zhang, Shang Gao, Yonghang Tai, and Jun Zhang.
\newblock Deep neural-based vulnerability discovery demystified: data, model
  and performance.
\newblock {\em Neural Computing and Applications}, 33(20):13287--13300, October
  2021.

\bibitem{codexglue}
Shuai Lu, Daya Guo, Shuo Ren, Junjie Huang, Alexey Svyatkovskiy, Ambrosio
  Blanco, Colin~B. Clement, Dawn Drain, Daxin Jiang, Duyu Tang, Ge~Li, Lidong
  Zhou, Linjun Shou, Long Zhou, Michele Tufano, Ming Gong, Ming Zhou, Nan Duan,
  Neel Sundaresan, Shao~Kun Deng, Shengyu Fu, and Shujie Liu.
\newblock {CodeXGLUE}: {A} machine learning benchmark dataset for code
  understanding and generation.
\newblock {\em CoRR}, abs/2102.04664, 2021.

\bibitem{mazuera-rozo_shallow_2021}
Alejandro Mazuera-Rozo, Anamaria Mojica-Hanke, Mario Linares-Vásquez, and
  Gabriele Bavota.
\newblock Shallow or {Deep}? {An} {Empirical} {Study} on {Detecting}
  {Vulnerabilities} using {Deep} {Learning}.
\newblock In {\em 2021 {IEEE}/{ACM} 29th {International} {Conference} on
  {Program} {Comprehension} ({ICPC})}, pages 276--287, May 2021.
\newblock ISSN: 2643-7171.

\bibitem{cyclomatic}
T.J. McCabe.
\newblock A complexity measure.
\newblock {\em IEEE Transactions on Software Engineering}, SE-2(4):308--320,
  1976.

\bibitem{alphacode}
Deep Mind.
\newblock Competitive programming with alphacode, 2022.

\bibitem{regvd}
Van-Anh Nguyen, Dai~Quoc Nguyen, Van Nguyen, Trung Le, Quan~Hung Tran, and Dinh
  Phung.
\newblock {ReGVD}: Revisiting graph neural networks for vulnerability
  detection.
\newblock In {\em 2022 IEEE/ACM 44th International Conference on Software
  Engineering: Companion Proceedings (ICSE-Companion)}, pages 178--182, 2022.

\bibitem{cotext}
Long Phan, Hieu Tran, Daniel Le, Hieu Nguyen, James Anibal, Alec Peltekian, and
  Yanfang Ye.
\newblock {CoTexT}: Multi-task learning with code-text transformer, 2021.

\bibitem{LIME_2016}
Marco~Tulio Ribeiro, Sameer Singh, and Carlos Guestrin.
\newblock {``Why Should I Trust You?''}: Explaining the predictions of any
  classifier.
\newblock KDD '16, page 1135–1144, New York, NY, USA, 2016. Association for
  Computing Machinery.

\bibitem{draper}
Rebecca Russell, Louis Kim, Lei Hamilton, Tomo Lazovich, Jacob Harer, Onur
  Ozdemir, Paul Ellingwood, and Marc McConley.
\newblock Automated {Vulnerability} {Detection} in {Source} {Code} {Using}
  {Deep} {Representation} {Learning}.
\newblock {\em Proceedings - 17th IEEE International Conference on Machine
  Learning and Applications, ICMLA 2018}, pages 757--762, 2019.
\newblock Publisher: IEEE.

\bibitem{hgvul}
Zihua Song, Junfeng Wang, Shengli Liu, Zhiyang Fang, Kaiyuan Yang, and
  Gu~Zhaoquan.
\newblock {HGVul}: A code vulnerability detection method based on heterogeneous
  source-level intermediate representation.
\newblock {\em Sec. and Commun. Netw.}, 2022, jan 2022.

\bibitem{typestate_analysis}
Robert~E. Strom and Shaula Yemini.
\newblock Typestate: A programming language concept for enhancing software
  reliability.
\newblock {\em IEEE Transactions on Software Engineering}, SE-12(1):157--171,
  1986.

\bibitem{Integrated_Gradients_2017}
Mukund Sundararajan, Ankur Taly, and Qiqi Yan.
\newblock Axiomatic attribution for deep networks.
\newblock In {\em Proceedings of the 34th International Conference on Machine
  Learning - Volume 70}, ICML'17, page 3319–3328. JMLR.org, 2017.

\bibitem{tang_comparative_2020}
Gaigai Tang, Lianxiao Meng, Huiqiang Wang, Shuangyin Ren, Qiang Wang, Lin Yang,
  and Weipeng Cao.
\newblock A {Comparative} {Study} of {Neural} {Network} {Techniques} for
  {Automatic} {Software} {Vulnerability} {Detection}.
\newblock In {\em 2020 {International} {Symposium} on {Theoretical} {Aspects}
  of {Software} {Engineering} ({TASE})}, pages 1--8, December 2020.

\bibitem{LIT_2020}
Ian Tenney, James Wexler, Jasmijn Bastings, Tolga Bolukbasi, Andy Coenen,
  Sebastian Gehrmann, Ellen Jiang, Mahima Pushkarna, Carey Radebaugh, Emily
  Reif, and Ann Yuan.
\newblock The language interpretability tool: Extensible, interactive
  visualizations and analysis for {NLP} models.
\newblock In {\em Proceedings of the 2020 Conference on Empirical Methods in
  Natural Language Processing: System Demonstrations}, pages 107--118, Online,
  October 2020. Association for Computational Linguistics.

\bibitem{tripp_taj_2009}
Omer Tripp, Marco Pistoia, Stephen~J. Fink, Manu Sridharan, and Omri Weisman.
\newblock {TAJ}: effective taint analysis of web applications.
\newblock {\em ACM SIGPLAN Notices}, 44(6):87--97, June 2009.

\bibitem{syncobert}
Xin Wang, Yasheng Wang, Fei Mi, Pingyi Zhou, Yao Wan, Xiao Liu, Li~Li, Hao Wu,
  Jin Liu, and Xin Jiang.
\newblock {SynCoBERT}: {Syntax}-{Guided} {Multi}-{Modal} {Contrastive}
  {Pre}-{Training} for {Code} {Representation}, September 2021.
\newblock arXiv:2108.04556 [cs].

\bibitem{vulcnn}
Yueming Wu, Deqing Zou, Shihan Dou, Wei Yang, Duo Xu, and Hai Jin.
\newblock {VulCNN}: an image-inspired scalable vulnerability detection system.
\newblock In {\em Proceedings of the 44th {International} {Conference} on
  {Software} {Engineering}}, {ICSE} '22, pages 2365--2376, New York, NY, USA,
  May 2022. Association for Computing Machinery.

\bibitem{acgdp}
Jiaxi Xu, Jun Ai, Jingyu Liu, and Tao Shi.
\newblock {ACGDP}: {An} {Augmented} {Code} {Graph}-{Based} {System} for
  {Software} {Defect} {Prediction}.
\newblock {\em IEEE Transactions on Reliability}, pages 1--10, 2022.
\newblock Conference Name: IEEE Transactions on Reliability.

\bibitem{GNNExplainer}
Rex Ying, Dylan Bourgeois, Jiaxuan You, Marinka Zitnik, and Jure Leskovec.
\newblock {GNN} explainer: {A} tool for post-hoc explanation of graph neural
  networks.
\newblock {\em CoRR}, abs/1903.03894, 2019.

\bibitem{astnn}
Jian Zhang, Xu~Wang, Hongyu Zhang, Hailong Sun, Kaixuan Wang, and Xudong Liu.
\newblock A {Novel} {Neural} {Source} {Code} {Representation} {Based} on
  {Abstract} {Syntax} {Tree}.
\newblock In {\em 2019 {IEEE}/{ACM} 41st {International} {Conference} on
  {Software} {Engineering} ({ICSE})}, pages 783--794, Montreal, QC, Canada, May
  2019. IEEE.

\bibitem{d2a_paper}
Yunhui Zheng, Saurabh Pujar, Burn Lewis, Luca Buratti, Edward Epstein, Bo~Yang,
  Jim Laredo, Alessandro Morari, and Zhong Su.
\newblock D2a: A dataset built for ai-based vulnerability detection methods
  using differential analysis.
\newblock In {\em Proceedings of the ACM/IEEE 43rd International Conference on
  Software Engineering: Software Engineering in Practice}, ICSE-SEIP '21, New
  York, NY, USA, 2021. Association for Computing Machinery.

\bibitem{zhou_devign_2019}
Yaqin Zhou, Shangqing Liu, Jingkai Siow, Xiaoning Du, and Yang Liu.
\newblock Devign: {Effective} vulnerability identification by learning
  comprehensive program semantics via graph neural networks.
\newblock {\em Advances in Neural Information Processing Systems}, 32:1--11,
  2019.

\end{thebibliography}



\end{document}